# Spatially varying coefficient modeling for large datasets: Eliminating $N$ from spatial regressions

Daisuke Murakami[1], Daniel A. Griffith[2]

[1]Department of Data Science, Institute of Statistical Mathematics,

10-3 Midori-cho, Tachikawa, Tokyo, 190-8562, Japan

Email: dmuraka@ism.ac.jp

[2]School of Economic, Political and Policy Science, The University of Texas, Dallas,

800 W Campbell Rd, Richardson, TX, 75080, USA

Email: dagriffith@utdallas.edu

**Abstract**: While spatially varying coefficient (SVC) modeling is popular in applied science, its computational burden is substantial. This is especially true if a multiscale property of SVC is considered. Given this background, this study develops a Moran's eigenvector-based spatially varying coefficients (M-SVC) modeling approach that estimates multiscale SVCs computationally efficiently. This estimation is accelerated through a (i) rank reduction, (ii) pre-compression, and (iii) sequential likelihood maximization. Steps (i) and (ii) eliminate the sample size $N$ from the likelihood function; after these steps, the likelihood maximization cost is independent of $N$. Step (iii) further accelerates the likelihood maximization so that multiscale SVCs can be estimated even if the number of SVCs, $K$, is large. The M-SVC approach is compared with geographically weighted regression (GWR) through Monte Carlo simulation experiments. These simulation results show that our approach is far faster than GWR when $N$ is large, despite numerically estimating $2K$ parameters while GWR numerically estimates only 1 parameter. Then, the proposed approach is applied to a land price analysis as an illustration. The developed SVC estimation approach is implemented in the R package "spmoran"

# 1. Introduction

Spatial and spatiotemporal data are increasing in size, and statistics for large spatial and spatiotemporal data are increasingly important. Unfortunately, spatial statistical approaches, which have been proposed in geostatistics (e.g., Cressie, 1993) and spatial econometrics (e.g., LeSage and Pace, 2009), typically require an inversion of a dense covariance matrix whose computational complexity is $O(N^3)$, where $N$ is the sample size. These approaches are not suitable for large samples.

To overcome this limitation, large spatial data modeling has been rapidly developing. In geostatistics, sparse and low rank approaches are representative of this development; the sparse approach includes covariance tapering (Furrer et al., 2006), use of stochastic partial differential equations (Lindgren et al., 2011), and the nearest neighbor Gaussian process (Datta et al., 2016), whereas the low rank approach includes fixed rank kriging (Cressie and Johannesson, 2008), the predictive process modeling (Banerjee et al., 2008), and the multiresolution approximation (Nychka et al., 2015) (see Heaton et al., 2016 for a review). These approaches are mainly for spatial interpolation. Meanwhile, spatial econometricians have studied large data modeling mainly for regression problems in the presence of spatial dependence. For example, polynomial-based approximations (Smirnov and Anselin, 2001; Griffith, 2004), a composite likelihood approximation (Arbia, 2014), and other likelihood approximations (e.g., LeSage and Pace, 2007) have been proposed. Burden et al. (2015) proposed a low rank approach. Spatial two-stage least squares (Kelejian and Prucha, 1998) is known as a computationally efficient approach.

Although spatial dependence and spatial heterogeneity are two general properties of spatial data (Tobler, 1970; Goodchild, 2004; Anselin, 2010), these preceding studies have paid attention to the former, whereas the latter has been somewhat overlooked. Nevertheless, consideration of spatial heterogeneity is crucially important to model spatial data appropriately (see, Fotheringham et al., 2002; Geniaux and Martinetti, 2017; Debarsy and Yang, 2018).

Given this background, this study focuses on spatially varying coefficient (SVC) modeling, which is a representative approach for spatial heterogeneity modeling. A large number of studies have examined SVC modeling (see Section 2). Recently, Fotheringham et al. (2017) and Murakami et al. (2018) revealed the importance of considering a multiscale property of SVCs to estimate them accurately; for example, the fist SVC might have a larger-scale map pattern than the other SVCs. However, if such a multiscale property is considered, the computational cost rapidly increases depending on not only $N$ but also the number of SVCs (see Section 2). How to estimate multiscale SVCs computationally efficiently remains unclear.

The objective of this study is to overcome this limitation. Specifically, we develop an approach to estimate multiscale SVCs computationally efficiently by extending the Moran's eigenvector-spatial filtering approach (Griffith 2003; Murakami et al. 2017). The subsequent sections are organized as follows. Section 2 briefly reviews existing SVC modeling approaches, and compares their computational properties. Section 3 introduces our SVC model, and Section 4 develops a fast SVC estimation method. Section 5 compares the computational time of the proposed approach with exiting approaches. Section 6 presents an empirical application. Finally, Section 7 concludes our discussion.

## 2. SVC modeling approaches

Suppose that response variables $y(s_i)$ are sampled at $N$ sample sites $s_i | i \in \{1, \cdots, N\}$, distributed in a study region, $D \subset R^2$. The linear SVC model is formulated as follows:

$$y(s_i) = \sum_{k=1}^{K} x_k(s_i)\beta_k(s_i) + \varepsilon(s_i), \qquad E[\varepsilon(s_i)] = 0, \qquad Var[\varepsilon(s_i)] = \sigma^2, \qquad (1)$$

where, $x_k(s_i)$ represents the $k$-th explanatory variable, $\beta_k(s_i)$ represents the $k$-th SVC, and $\sigma^2$ is a variance parameter. SVC modeling aims to estimate and infer about $\beta_k(s_i)$.

Geographical weighted regression (GWR; Brunsdon et al., 1996) is the representative approach for SVC modeling. GWR estimates $\beta_k(s_i)$ by assigning more weights on samples nearby than those far from the site $s_i$. The estimator for $\boldsymbol{\beta}(s_i) = [\beta_1(s_i), \cdots \beta_K(s_i)]'$, where " ′ " denotes the matrix transpose operator, is given as

$$\hat{\boldsymbol{\beta}}(s_i) = [\mathbf{X}'\mathbf{G}(s_i)\mathbf{X}]^{-1}\mathbf{X}'\mathbf{G}(s_i)\mathbf{y} \qquad (2)$$

where $\mathbf{X}$ is an $N \times K$ matrix of explanatory variables, $\mathbf{y}$ is an $N \times 1$ vector of response variables, and $\mathbf{G}(s_i)$ is an $N \times N$ diagonal matrix whose $j$-th element $g(s_i, s_j)$ represents the weight assigned to the $j$-th sample. $g(s_i, s_j)$ is given by a distance decay kernel. For example, Wheeler and Calder (2007) and Wheeler and Waller (2009) used the exponential kernel, $g(s_i, s_j) = \exp(-d(s_i, s_j)/b)$, where $b$ is the bandwidth determining the scale of the SVCs; $\hat{\boldsymbol{\beta}}(s_i)$ have a large-scale map pattern when $b$ is large, and a small-scale pattern when $b$ is small. The bandwidth is estimated by minimizing the corrected Akaike Information Criterion (AICc), or minimizing the predictive error from cross-validation (CV). GWR has widely been applied in environmental studies (e.g., Harris et al., 2010A; Dong et al., 2018), criminology (e.g., Cahill and Mulligan, 2007), epidemiology (e.g., Nakaya et al., 2005), and many other fields.

The single bandwidth implicitly imposes a strong assumption that all the SVCs have the same scale of spatial variation. However, some coefficients might have a larger scale pattern than the others. To consider such a multiscale property, Yang (2014) and Fotheringham et al. (2017) proposed the flexible-band GWR (FB-GWR; also called the multiscale GWR), which replaces the common $b$ with the SVC-specific bandwidth, $b_k$. The FB-GWR is estimated by a back-fitting algorithm that sequentially iterates the AICc maximization (or the CV) for each $b_k$ until convergence. Recently, Lu et al. (2018) proposed a computationally efficient algorithm for FB-GWR model estimation, achieving more than a 60 % reduction in computational time. Yet, because the original FB-GWR is slow even when $N \leq 5{,}000$, as we show later, a much faster algorithm might be required for larger samples.

The Bayesian SVC model (B-SVC model; Gelfand et al., 2003) furnishes another approach. This model assumes that the $k$-th SVC, $\boldsymbol{\beta}_k = [\beta_k(s_1), \cdots \beta_k(s_N)]'$, obeys the following Gaussian process:

$$\boldsymbol{\beta}_k \sim N\big(b_k \mathbf{1}, \tau_k \mathbf{C}(r_k)\big), \tag{3}$$

where $b_k$, $\tau_k$, and $r_k$ are parameters, and $\mathbf{1}$ ($N \times 1$) is a vector of ones. $\mathbf{C}(r_k)$ ($N \times N$) is a correlation matrix whose ($i$, $j$)-th element, $c_{ij}(r_k)$, is given by a distance decay kernel to capture spatial dependence, such as the exponential kernel, $c_{ij}(r_k)$=exp[-$d(\underline{s}_i, s_j)/r_k$], where $d(\underline{s}_i, s_j)$ is the Euclidean distance between sites $s_i$ and $s_j$. The B-SVC model, Eqs. (1) and (2), is estimated by Markov chain Monte Carlo (MCMC) techniques. Wheeler and Waller (2009), and Finley et al. (2011A), among others, have compared the B-SVC modeling with GWR, and showed the robustness of the B-SVC approach.

Unfortunately, the computational cost for B-SVC modeling is disappointedly heavy when $N$ and/or $K$ is large (Finley et al., 2011A). This is because the MCMC simulation requires inverting the dense matrices $\mathbf{C}(r_1),\dots\mathbf{C}(r_K)$ in each iteration. To lighten this cost, Finley et al. (2011B) applied a rank reduction to the B-SVC modeling, coded in C++ and Fortran. However, their MCMC simulation with 50,000 iterations still requires ~72 hours when $N = 8,774$ and $K = 5$. A faster approach that works within minutes even with larger $N$ and $K$ is more preferable in practice.

A Moran eigenvector SVC (M-SVC) modeling approach (Griffith, 2008), which is an extension of the Moran eigenvector spatial filtering approach (see Griffith 2003; Griffith and Chun, 2014), furnishes a third option for multiscale SVC modeling. While instability of this approach was alluded to by Helbich and Griffith (2016) and Oshan and Fotheringham (2018), Murakami et al. (2017) extended it to approximate the B-SVC model, and Murakami et al. (2017; 2018) demonstrated its estimation accuracy through Monte Carlo experiments. Although their approach does not require MCMC iterations, unlike the B-SVC modeling approach, it is still slow when $N$ is large because it requires an eigen-decomposition (complexity: $O(N^3)$) and numerical estimation of $2K$ parameters.

Thus, how to estimate multiscale SVCs computationally efficiently remains elusive. Because SVC modeling is popular in a wide variety of applied fields, achieving fast computation without relying on high performance computing environment is desirable.

# 3. Moran eigenvector SVC (M-SVC) modeling

## 3.1. Model

This approach is based on the Moran coefficient (MC; Moran, 1950), which is a diagnostic statistic of spatial dependence. MC for $\mathbf{y} = [y(s_1),\ldots y(s_N)]'$ is formulated as:

$$MC[\mathbf{y}] = \frac{N}{\mathbf{1}'\mathbf{C}\mathbf{1}} \frac{\mathbf{y}'(\mathbf{I} - \mathbf{11}'/N)\mathbf{C}(\mathbf{I} - \mathbf{11}'/N)\mathbf{y}}{\mathbf{y}'(\mathbf{I} - \mathbf{11}'/N)\mathbf{y}}, \qquad (4)$$

where $\mathbf{C}$ is a symmetric spatial proximity matrix with zero diagonal entries, and $\mathbf{I} - \mathbf{11}'/N$ is a centering matrix. $MC[\mathbf{y}] > -\frac{1}{N-1}$ if $y(s_1),\ldots y(s_N)$ are positively spatially dependent, and $MC[\mathbf{y}] < -\frac{1}{N-1}$ if they are negatively dependent (Griffith, 2003). Following Drey et al. (2006) and Murakami and Griffith (2015), we define $\mathbf{C}$ by a matrix whose $(i,j)$-th element is $\exp\left(-d\left(s_i, s_j\right)/r\right)$, where the range parameter $r$ is given by the maximum distance in the minimum spanning tree connecting all sample sites. Instead of the fixed $r$, we introduce another parameter to control the scale/range of spatial dependence. This replacement is needed for fast computation.

Suppose that $\mathbf{E} = [\mathbf{e}_1, ..., \mathbf{e}_L]$ ($N \times L$) is a matrix of the $L$ ($< N$) eigenvectors and $\mathbf{\Lambda}$ ($L \times L$) is a diagonal matrix of the corresponding $L$ eigenvalues $\{\lambda_1, ..., \lambda_L\}$ of the matrix $(\mathbf{I} - \mathbf{11}'/N)\mathbf{C}(\mathbf{I} - \mathbf{11}'/N)$. Then the following equation holds: $MC[\mathbf{e}_l] = (N/\mathbf{1}'\mathbf{C}\mathbf{1})\lambda_l$, which means that the eigenvectors corresponding to larger positive eigenvalues explain stronger spatial dependence, whereas the opposite is true for negative spatial dependence. Thus, the eigenvectors provide orthogonal map pattern descriptions of latent spatial dependence, with each magnitude being indexed by the MC (Griffith, 2003; Tiefelsdorf and Griffith, 2007). Because positive spatial dependence has implicitly been assumed in SVCs in existing studies (e.g., Fotheringham et al., 2002), this study applies $L$ eigen-pairs corresponding to positive eigenvalues from $\mathbf{E}$ and $\mathbf{\Lambda}$.

The M-SVC model of Murakami et al. (2017) is formulated as follows:

$$\mathbf{y} = b_1\mathbf{1} + \sum_{k=2}^{K} \mathbf{x}_k \circ \mathbf{\beta}_k + \mathbf{E}\mathbf{\gamma}_1 + \mathbf{\varepsilon}, \qquad \mathbf{\gamma}_1 \sim N(\mathbf{0}, \tau_1^2 \mathbf{\Lambda}^{\alpha_1}), \qquad \mathbf{\varepsilon} \sim N(\mathbf{0}, \sigma^2 \mathbf{I}), \qquad (5)$$

where " $\circ$ " represents the column-wise product operator, $\mathbf{y}$ ($N \times 1$) is a vector of response variables, and $\mathbf{x}_k$ ($N \times 1$) is a vector of $k$-th covariates. $\mathbf{E}\mathbf{\gamma}_1$ is the term to capture residual spatial dependence. Murakami and Griffith (2018) showed that the term $\mathbf{E}\mathbf{\gamma}_1$ with $L = 200$ greatly reduces residual spatial dependence.

The $k$-th SVC is parameterized as follows:

$$\mathbf{\beta}_k = b_k\mathbf{1} + \mathbf{E}\mathbf{\gamma}_k, \qquad \mathbf{\gamma}_k \sim N(\mathbf{0}, \tau_k^2 \mathbf{\Lambda}^{\alpha_k}), \qquad (6)$$

where $b_k$ is a coefficient. The SVC consists of the constant term $b_k\mathbf{1}$ and the spatially varying term $\mathbf{E}\mathbf{\gamma}_k$. The term $\mathbf{E}\mathbf{\gamma}_k$ has the following properties: (i) it is interpretable in terms of the MC; and, (ii) the eigenvectors have zero means. Property (i) is important to model spatial dependence variations accurately as discussed. Property (ii) is required to make the mean term, $b_k\mathbf{1}$, and $\mathbf{E}\boldsymbol{\gamma}_k$ identifiable (see, Gelfand et al., 2003).

Vector $\boldsymbol{\gamma}_k$ contains random coefficients that depend on two shrinkage parameters,

$\tau_k^2$ and $\alpha_k$. Parameter $\tau_k^2$ controls the variance of the spatial variation of $\boldsymbol{\beta}_k$. Parameter $\alpha_k$ controls the spatial scale; large $\alpha_k$ shrinks coefficients on minor eigen-pairs with small eigenvalues, and the resulting SVCs have a large-scale map pattern. The opposite is true for small $\alpha_k$. Thus, $\alpha_k$ estimates the scale of the SVCs instead of the range parameter,

which we consider as fixed. This replacement enables drastically accelerating model

estimation while maintaining flexibility in estimating the scale. Murakami and Griffith

(2018) showed that the estimation error is small when $\alpha_k$ rather than $r$ is estimated.

Given $\mathbf{x}_1 = \mathbf{1}$ and $\tilde{\mathbf{E}}_1 = (\mathbf{x}_1 \circ \mathbf{E}) = \mathbf{E}$, the M-SVC model is expressed as

$$\mathbf{y} = \mathbf{X}\mathbf{b} + \sum_{k=1}^{K} \tilde{\mathbf{E}}_k \mathbf{V}(\boldsymbol{\theta}_k)\mathbf{u}_k + \boldsymbol{\varepsilon} \qquad \mathbf{u}_k \sim N(\mathbf{0},\ \sigma^2\mathbf{I}) \qquad \boldsymbol{\varepsilon} \sim N(\mathbf{0},\ \sigma^2\mathbf{I}) \qquad (7)$$

where $\mathbf{X} = [\mathbf{x}_1,... \ \mathbf{x}_K]$, $\mathbf{b} = [b_1,... \ b_K]'$, $\boldsymbol{\theta}_k \in \{\tau_k,\ \alpha_k\}$, and $\mathbf{V}(\boldsymbol{\theta}_k) = (\tau_k/\sigma)\boldsymbol{\Lambda}^{\alpha_k/2}$. Eq.(7)

suggests that the M-SVC model is identical to the linear mixed effects model (e.g., Bates,

2010).

## 3.2. Parameter estimation

The M-SVC model is estimated by the Type II restricted likelihood (empirical

Bayes) method that maximizes $loglik_R(\boldsymbol{\Theta}) = \int loglik(\mathbf{y}|\mathbf{b}, \boldsymbol{\Theta})d\mathbf{b}$, where $\boldsymbol{\Theta} \in \{\boldsymbol{\theta}_1, \cdots \boldsymbol{\theta}_K\}$

and $loglik(\mathbf{y}|\mathbf{b}, \boldsymbol{\Theta}) = \int \cdots \int p(\mathbf{y}, \mathbf{u}_1, \cdots \mathbf{u}_K|\mathbf{b}, \boldsymbol{\Theta}) p(\mathbf{u}_1, \cdots \mathbf{u}_K) d\mathbf{u}_1 \cdots d\mathbf{u}_K$. This restricted

likelihood has an analytic expression if $p(\mathbf{y}, \mathbf{u}_1, \cdots \mathbf{u}_K|\mathbf{b}, \boldsymbol{\Theta})$ and $p(\mathbf{u}_1, \cdots \mathbf{u}_K)$ are

Gaussians probability density functions. Using this property, the restricted log-likelihood

for Eq.(7) is derived as (see Bates, 2010; Murakami et al., 2017):

$$loglik_R(\boldsymbol{\Theta}) = -\frac{1}{2}ln|\mathbf{P}| - \frac{N-K}{2}\left(1 + ln\left(\frac{2\pi d(\boldsymbol{\Theta})}{N-K}\right)\right), \tag{8}$$

$$\mathbf{P} = \begin{bmatrix} \mathbf{X}'\mathbf{X} & \mathbf{X}'\tilde{\mathbf{E}}_1\mathbf{V}(\boldsymbol{\theta}_1) & \cdots & \mathbf{X}'\tilde{\mathbf{E}}_K\mathbf{V}(\boldsymbol{\theta}_K) \\ \mathbf{V}(\boldsymbol{\theta}_1)\tilde{\mathbf{E}}'_1\mathbf{X} & \mathbf{V}(\boldsymbol{\theta}_1)\tilde{\mathbf{E}}'_1\tilde{\mathbf{E}}_1\mathbf{V}(\boldsymbol{\theta}_1) + \mathbf{I} & \cdots & \mathbf{V}(\boldsymbol{\theta}_1)\tilde{\mathbf{E}}'_1\tilde{\mathbf{E}}_K\mathbf{V}(\boldsymbol{\theta}_K) \\ \vdots & \vdots & \ddots & \vdots \\ \mathbf{V}(\boldsymbol{\theta}_K)\tilde{\mathbf{E}}'_K\mathbf{X} & \mathbf{V}(\boldsymbol{\theta}_K)\tilde{\mathbf{E}}'_K\tilde{\mathbf{E}}_1\mathbf{V}(\boldsymbol{\theta}_1) & \cdots & \mathbf{V}(\boldsymbol{\theta}_K)\tilde{\mathbf{E}}'_K\tilde{\mathbf{E}}_K\mathbf{V}(\boldsymbol{\theta}_K) + \mathbf{I} \end{bmatrix}. \tag{9}$$

The derivative $d(\boldsymbol{\Theta})$ equals

$$d(\boldsymbol{\Theta}) = \left\|\mathbf{y} - \mathbf{X}\hat{\mathbf{b}} - \sum_{k=1}^{K} \tilde{\mathbf{E}}_k\mathbf{V}(\boldsymbol{\theta}_k)\hat{\mathbf{u}}_k\right\|^2 + \sum_{k=1}^{K} \|\hat{\mathbf{u}}_k\|^2, \tag{10}$$

where

$$\begin{bmatrix} \hat{\mathbf{b}} \\ \hat{\mathbf{u}}_1 \\ \vdots \\ \hat{\mathbf{u}}_K \end{bmatrix} = \mathbf{P}^{-1} \begin{bmatrix} \mathbf{X}'\mathbf{y} \\ \mathbf{V}(\boldsymbol{\theta}_1)\tilde{\mathbf{E}}'_1\mathbf{y} \\ \vdots \\ \mathbf{V}(\boldsymbol{\theta}_K)\tilde{\mathbf{E}}'_K\mathbf{y} \end{bmatrix}. \tag{11}$$

Eq. (10) balances the trade-off between accuracy and complexity of the model. The

parameters are estimated by the following steps:

(i)      $\boldsymbol{\Theta} \in \{\boldsymbol{\theta}_1, \dots \boldsymbol{\theta}_K\}$ are estimated by maximizing Eq. (8)

(ii)      $\{\mathbf{b}, \mathbf{u}_1, \cdots \mathbf{u}_K\}$ are estimated by Eq. (11)

(iii)    The SVCs are estimated by $\widehat{\boldsymbol{\beta}}_k = \hat{b}_k \mathbf{1} + \mathbf{E}\hat{\boldsymbol{\gamma}}_k = \hat{b}_k \mathbf{1} + \mathbf{E}\mathbf{V}(\widehat{\boldsymbol{\theta}}_k)\widehat{\mathbf{u}}_k$.

# 4. Fast M-SVC modeling

Unfortunately, the Moran eigenvector approach is computationally inefficient in terms of both modeling and estimation. Sections 4.1 presents how to reduce the cost for modeling, whereas Section 4.2 presents how to reduce the cost for estimation.

## 4.1. Modeling

The cost for the eigen-decomposition of $(\mathbf{I} - \mathbf{11}'/N)\mathbf{C}(\mathbf{I} - \mathbf{11}'/N)$ is $O(N^3)$, which is intractable for large $N$. Besides, if $\mathbf{C}$ is given using a distance-kernel like in our case, the $N \times N$ matrix must be stored before the decomposition. The modeling is inefficient in terms of both computational complexity and memory usage.

To address these problems, we apply a Nystrom extension (Drineas and Mahoney, 2005) based approximation of Moran eigenvectors and eigenvalues (Murakami and Griffith 2018) that is formulated as follows:[1]

$$\widehat{\mathbf{E}} = \left[ \mathbf{C}_{NL} - \mathbf{1} \otimes \frac{\mathbf{1}'_L(\mathbf{C}_L + \mathbf{I}_L)}{L} \right] \mathbf{E}_L(\boldsymbol{\Lambda}_L + \mathbf{I}_L)^{-1} \tag{12}$$

$$\widehat{\boldsymbol{\Lambda}} = \frac{L + N}{L}(\boldsymbol{\Lambda}_L + \mathbf{I}_L) - \mathbf{I}_L \tag{13}$$

---

[1]  Griffith (2000, 2015) proposed other eigen-approximations for regular square tessellation data.

where "$\otimes$" denotes the Kronecker product operator. $\mathbf{C}_L$ is a $L \times L$ spatial proximity matrix with $L$ knots ($L \ll N$) distributed across the study region. The knots are given by the $k$-means cluster centers calculate using the spatial coordinates of sample sizes (see Zhang and Kwok, 2010). $\mathbf{C}_{NL}$ ($N \times L$) is a proximity matrix between the $L$ knots and the $N$ sample sites. The set $\{\mathbf{E}_L, \mathbf{\Lambda}_L, \mathbf{I}_L, \mathbf{1}_L\}$ is defined just like the set $\{\mathbf{E}, \mathbf{\Lambda}, \mathbf{I}, \mathbf{1}\}$ for the knots.

Following Murakami and Griffith (2018), which showed that 200 eigen-pairs are sufficient to model spatial dependence, $L$ is fixed by the number of positive eigenvalues in $\widehat{\mathbf{\Lambda}}$ if it is below 200, and 200 otherwise. Given that $L$ as fixed, the computational complexity for the eigen-approximation yields $O(N)$, whereas the memory usage yields $N \times L$. Regarding the number of SVCs, considering the difficulty of identifying them (see Wheeler and Tiefelsdorf, 2005; Farber and Páez, 2007; Helbich and Griffith, 2016), among others, at most 10 SVCs might be reasonable.

Thus, we assume that $L$ is fixed and $K \ll N$. When a large number of explanatory variables must be considered, SVCs can be given only for some focused/selected explanatory variables, whereas constant coefficients are assumed for the others. The $k$-th SVC is substituted by a constant coefficient by replacing $\mathbf{\beta}_k = b_k \mathbf{1} + \mathbf{E}\mathbf{\gamma}_k$ with $b_k \mathbf{1}$. This property is useful to avoid increasing $K$ and assuring the identifiability of SVCs.

Our approximate M-SVC model is formulated using Eqs. (14) and (15):

$$\mathbf{y} = b_1 \mathbf{1} + \sum_{k=2}^{K} \mathbf{x}_k {}^{\circ} \boldsymbol{\beta}_k + \hat{\mathbf{E}} \boldsymbol{\gamma}_1 + \boldsymbol{\varepsilon}, \qquad \boldsymbol{\gamma}_1 {\sim} N\big(\mathbf{0},\ \tau_1^2 \hat{\boldsymbol{\Lambda}}^{\alpha_1}\big), \qquad \boldsymbol{\varepsilon} {\sim} N\big(\mathbf{0},\ \sigma^2 \mathbf{I}\big), \quad (14)$$

$$\boldsymbol{\beta}_k = b_k \mathbf{1} + \hat{\mathbf{E}} \boldsymbol{\gamma}_k, \qquad \boldsymbol{\gamma}_k {\sim} N\big(\mathbf{0},\ \tau_k^2 \hat{\boldsymbol{\Lambda}}^{\alpha_k}\big), \qquad\qquad (15)$$

The next section explains how to estimate the model computationally efficiently.

## 4.2. Estimation

The likelihood maximization can be very slow because it includes $2K$ shrinkage parameters in $\boldsymbol{\Theta} \in \{\boldsymbol{\theta}_1, \dots \boldsymbol{\theta}_K\}$, where $\boldsymbol{\theta}_k \in \{\tau_k, \alpha_k\}$, that do not have closed form solutions. For example, if 10 SVCs are assumed, 20 shrinkage parameters must be estimated numerically. The M-SVC model estimation is much more difficult than GWR, which numerically estimates 1 bandwidth, or FB-GWR, which estimates $K$ bandwidths.

This section shows that the computational cost for the M-SVC model estimation is drastically reduced by employing selected matrix tricks. Section 4.2.1 eliminates $N$ from the likelihood function, and Section 4.2.2 derives a sequential approach to maximize the likelihood.

### 4.2.1. Elimination of "$N$" from the likelihood function

To estimate $\boldsymbol{\Theta} \in \{\boldsymbol{\theta}_1, \dots \boldsymbol{\theta}_K\}$, the restricted log-likelihood must be evaluated repeatedly. To reduce this burden, this study eliminates the matrices and vectors whose size depends on $N$, before the estimation. Specifically, the likelihood is evaluated is the following steps:

(a)    $\mathbf{M}_{0,0} = \mathbf{X}'\mathbf{X}$,  $\mathbf{M}_{0,k} = \mathbf{X}'(\mathbf{x}_k{}^\circ\mathbf{E})$,  $\mathbf{M}_{k,\bar{k}} = (\mathbf{x}_k{}^\circ\mathbf{E})'(\mathbf{x}_{\bar{k}}{}^\circ\mathbf{E})$,  $\mathbf{m}_0 = \mathbf{X}'\mathbf{y}$,  $\mathbf{m}_k = (\mathbf{x}_k{}^\circ\mathbf{E})'\mathbf{y}$,

and  $m_{y,y} = \mathbf{y}'\mathbf{y}$  are evaluated

(b)    The restricted log-likelihood is rewritten by substituting the matrices and vectors into Eq.(8) as follows:

$$loglik_R(\boldsymbol{\Theta}) = -\frac{1}{2}ln|\mathbf{P}| - \frac{N-K}{2}\left(1 + ln\left(\frac{2\pi}{N-K}\boldsymbol{d}(\boldsymbol{\Theta})\right)\right), \tag{16}$$

where

$$\mathbf{P} = \begin{bmatrix} \mathbf{M}_{0,0} & \mathbf{M}_{0,1}\mathbf{V}(\boldsymbol{\theta}_1) & \cdots & \mathbf{M}_{0,K}\mathbf{V}(\boldsymbol{\theta}_K) \\ \mathbf{V}(\boldsymbol{\theta}_1)\mathbf{M}_{1,0} & \mathbf{V}(\boldsymbol{\theta}_1)\mathbf{M}_{1,1}\mathbf{V}(\boldsymbol{\theta}_1)+\mathbf{I} & \cdots & \mathbf{V}(\boldsymbol{\theta}_1)\mathbf{M}_{1,K}\mathbf{V}(\boldsymbol{\theta}_K) \\ \vdots & \vdots & \ddots & \vdots \\ \mathbf{V}(\boldsymbol{\theta}_K)\mathbf{M}_{K,0} & \mathbf{V}(\boldsymbol{\theta}_K)\mathbf{M}_{K,1}\mathbf{V}(\boldsymbol{\theta}_1) & \cdots & \mathbf{V}(\boldsymbol{\theta}_K)\mathbf{M}_{K,K}\mathbf{V}(\boldsymbol{\theta}_K)+\mathbf{I} \end{bmatrix}, \tag{17}$$

and  $\boldsymbol{d}(\boldsymbol{\Theta}) = \|\hat{\boldsymbol{\varepsilon}}\|^2 + \sum_{k=1}^{K}\|\hat{\mathbf{u}}_k\|^2$,  where

$$\|\hat{\boldsymbol{\varepsilon}}\|^2 = m_{y,y} - 2[\hat{\mathbf{b}}', \hat{\mathbf{u}}'_1, \cdots \hat{\mathbf{u}}'_K]\begin{bmatrix} \mathbf{m}_0 \\ \mathbf{V}(\boldsymbol{\theta}_1)\mathbf{m}_1 \\ \vdots \\ \mathbf{V}(\boldsymbol{\theta}_K)\mathbf{m}_K \end{bmatrix} + [\hat{\mathbf{b}}', \hat{\mathbf{u}}'_1, \cdots \hat{\mathbf{u}}'_K]\mathbf{P}_0\begin{bmatrix} \hat{\mathbf{b}} \\ \hat{\mathbf{u}}_1 \\ \vdots \\ \hat{\mathbf{u}}_K \end{bmatrix}, \tag{18}$$

$$\begin{bmatrix} \hat{\mathbf{b}} \\ \hat{\mathbf{u}}_1 \\ \vdots \\ \hat{\mathbf{u}}_K \end{bmatrix} = \mathbf{P}^{-1}\begin{bmatrix} \mathbf{m}_0 \\ \mathbf{V}(\boldsymbol{\theta}_1)\mathbf{m}_1 \\ \vdots \\ \mathbf{V}(\boldsymbol{\theta}_K)\mathbf{m}_K \end{bmatrix}. \tag{19}$$

$\mathbf{P}_0 = \mathbf{P}$, in which  $\mathbf{V}(\boldsymbol{\theta}_k)\mathbf{M}_{k,k}\mathbf{V}(\boldsymbol{\theta}_k)+\mathbf{I}$  is replaced with  $\mathbf{V}(\boldsymbol{\theta}_k)\mathbf{M}_{k,k}\mathbf{V}(\boldsymbol{\theta}_k)$.

(c)    $\boldsymbol{\Theta}$ is estimated by maximizing Eq.(16)

Because Eq. (16) does not include any matrix whose size depends on $N$, the computational complexity for the estimation of $\boldsymbol{\Theta}$ is independent of $N$. Yet, the maximization in step (c) is too slow if $K$ is large. To reduce this computational cost, the next section develops an approach to estimate each $\boldsymbol{\theta}_k$ sequentially.

### 4.2.2. Sequential likelihood maximization

This section discusses maximizing $loglik_R(\boldsymbol{\Theta})$ by sequentially solving Eq.(20) for each $k$ until the likelihood value converges such that

$$\widehat{\boldsymbol{\theta}}_k = \underset{\boldsymbol{\theta}_k}{\operatorname{argmax}} \; loglik_R(\boldsymbol{\theta}_k|\boldsymbol{\Theta}_{-k}), \tag{20}$$

where $\boldsymbol{\Theta}_{-k} \in \{\boldsymbol{\theta}_1,... \boldsymbol{\theta}_{k-1}, \boldsymbol{\theta}_{k+1},... \boldsymbol{\theta}_K\}$. This maximization requires evaluations of $\mathbf{P}^{-1}$ and $|\mathbf{P}|$, both of whose complexities are $O((K + KL)^3)$; complexity increases rapidly as $K$ grows, while $L$ is constrained to be at most 200. Because $loglik_R(\boldsymbol{\theta}_k|\boldsymbol{\Theta}_{-k})$ must be evaluated many times to maximize it, the computational burden is non-ignorable.

To reduce the computational cost, Sections 4.2.2.1 and 4.2.2.2 present fast approaches to evaluate $\mathbf{P}^{-1}$ and $|\mathbf{P}|$, respectively. Although these sections assume estimation of $\boldsymbol{\theta}_K$ as an example, the same approach is available to estimate $\{\boldsymbol{\theta}_1,... \boldsymbol{\theta}_{K-1}\}$.

### 4.2.2.1. Fast evaluation of the term including $\mathbf{P}^{-1}$

Suppose that $\mathbf{m}_{-K} = [\mathbf{m'}_0 \quad \mathbf{m'}_1 \quad \cdots \quad \mathbf{m'}_{K-1}]'$. Then, as derived in Appendix.

1, Eq. (19), including the inversion $\mathbf{P}^{-1}$, can be expressed as follows:

$$
\begin{bmatrix} \widehat{\mathbf{b}} \\ \widehat{\mathbf{u}}_1 \\ \vdots \\ \widehat{\mathbf{u}}_K \end{bmatrix} = \begin{bmatrix} \widetilde{\mathbf{V}}_{-K}^{-1} & \mathbf{O} \\ \mathbf{O} & \mathbf{V}(\boldsymbol{\theta}_K)^{-1} \end{bmatrix} \mathbf{Q}^{-1} \begin{bmatrix} \mathbf{m}_{-K} \\ \mathbf{m}_K \end{bmatrix} \tag{21}
$$

$$
- \begin{bmatrix} \widetilde{\mathbf{V}}_{-K}^{-1} \mathbf{Q}^*_{-K,K} \\ \mathbf{V}(\boldsymbol{\theta}_K)^{-1} \mathbf{Q}^*_{K,K} \end{bmatrix} \left( \mathbf{V}(\boldsymbol{\theta}_K)^2 + \mathbf{Q}^*_{K,K} \right)^{-1} \left[ \mathbf{Q}^*_{K,-K} \mathbf{m}_{-K} + \mathbf{Q}^*_{K,K} \mathbf{m}_K \right],
$$

where $\widetilde{\mathbf{V}}_{-K} = \begin{bmatrix} \mathbf{I} & & & \\ & \mathbf{V}_1 & & \\ & & \ddots & \\ & & & \mathbf{V}_{K-1} \end{bmatrix}$. $\mathbf{Q} = \begin{bmatrix} \widetilde{\mathbf{M}}_{-K,-K} + \widetilde{\mathbf{V}}_{-K}^{-2} & \widetilde{\mathbf{M}}_{-K,K} \\ \widetilde{\mathbf{M}}_{K,-K} & \mathbf{M}_{K,K} \end{bmatrix}$, where $\widetilde{\mathbf{M}}_{-K,-K} =$

$\begin{bmatrix} \mathbf{M}_{0,0} & \mathbf{M}_{0,1} & \cdots & \mathbf{M}_{0,K-1} \\ \mathbf{M}_{1,0} & \mathbf{M}_{1,1} + \mathbf{V}_1^{-2} & \cdots & \mathbf{M}_{1,K-1} \\ \vdots & \vdots & \ddots & \vdots \\ \mathbf{M}_{K-1,0} & \mathbf{M}_{K-1,1} & \cdots & \mathbf{M}_{K-1,K-1} + \mathbf{V}_{K-1}^{-2} \end{bmatrix}$ and $\widetilde{\mathbf{M}}_{-K,K} = [\mathbf{M}_{K,0} \quad \mathbf{M}_{K,1} \quad \cdots \quad \mathbf{M}_{K,K-1}]'$,

and $\begin{bmatrix} \mathbf{Q}^*_{-K,-K} & \mathbf{Q}^*_{-K,K} \\ \mathbf{Q}^*_{K,-K} & \mathbf{Q}^*_{K,K} \end{bmatrix} = \mathbf{Q}^{-1}$.

Different from Eq. (19), Eq. (21) does not include the inversion of a large matrix including $\boldsymbol{\theta}_K$, which we want to estimate. Eq. (21) still has $\mathbf{Q}^{-1}$, whose complexity is $O((K + KL)^3)$. However, because the $\mathbf{Q}$ matrix does not include $\boldsymbol{\theta}_K$, if only $\mathbf{Q}^{-1}$ is evaluated one time, it can be fixed in the iterative calculation to maximize $loglik_R(\boldsymbol{\theta}_k | \boldsymbol{\Theta}_{-k})$. The most computationally demanding part in the iterative calculation is $\left( \mathbf{V}(\boldsymbol{\theta}_K)^2 + \mathbf{Q}^*_{K,K} \right)^{-1}$ whose computational cost is $O(L^3)$, which is trivial under our constraint of $L \leq 200$.

### 4.2.2.2. Fast evaluation of $|\mathbf{P}|$

As derived in Appendix. 2, $|\mathbf{P}|$ can be expanded as follows:

$$|\mathbf{P}| = \left|\widetilde{\mathbf{V}}_{-K}\right|^2 |\mathbf{V}(\boldsymbol{\theta}_K)|^2 \left|\widetilde{\mathbf{M}}_{-K,-K} + \widetilde{\mathbf{V}}_{-K}^{-2}\right| \Big|\mathbf{V}(\boldsymbol{\theta}_K)^{-2} + \mathbf{M}_{K,K}$$

$$- \widetilde{\mathbf{M}}_{K,-K}(\widetilde{\mathbf{M}}_{-K,-K} + \widetilde{\mathbf{V}}_{-K}^{-2})^{-1}\widetilde{\mathbf{M}}_{-K,K}\Big|. \tag{22}$$

The computations of $\left|\widetilde{\mathbf{V}}_{-K}\right|^2$ and $|\mathbf{V}(\boldsymbol{\theta}_K)|^2$ are immediate because $\widetilde{\mathbf{V}}_{-K}$ and $\mathbf{V}(\boldsymbol{\theta}_K)$ are diagonal matrices. The complexities for $\left|\widetilde{\mathbf{M}}_{-K,-K} + \widetilde{\mathbf{V}}_{-K}^{-2}\right|$ and $(\widetilde{\mathbf{M}}_{-K,-K} + \widetilde{\mathbf{V}}_{-K}^{-2})^{-1}$ are $O((K+(K\text{-}1)L)^3)$, which is relatively large. However, because these matrices do not include $\boldsymbol{\theta}_K$, they can be evaluated before the parameter estimation. After all, the most demanding calculation that must be iterated in the likelihood maximization is $\left|\mathbf{V}(\boldsymbol{\theta}_K)^{-2} + \widetilde{\mathbf{M}}_{K,K}\right|$, where $\widetilde{\mathbf{M}}_{K,K} = \mathbf{M}_{K,K} - \widetilde{\mathbf{M}}_{K,-K}(\widetilde{\mathbf{M}}_{-K,-K} + \widetilde{\mathbf{V}}_{-K}^{-2})^{-1}\widetilde{\mathbf{M}}_{-K,K}$ is given a priori. The complexity for this determinant evaluation is $O(L^3)$.

In summary, if matrices and vectors that do not include $\boldsymbol{\theta}_K$ are processed a priori, computational complexity for both $\left|\widetilde{\mathbf{M}}_{-K,-K} + \widetilde{\mathbf{V}}_{-K}^{-2}\right|$ and $(\widetilde{\mathbf{M}}_{-K,-K} + \widetilde{\mathbf{V}}_{-K}^{-2})^{-1}$, which are contained in the function $loglik_R(\boldsymbol{\theta}_k|\boldsymbol{\Theta}_{-k})$, becomes $O(L^3)$.

### 4.3. Summary

Our approach is summarized in Figure 1. In the modeling step, we apply (i) a rank reduction, and each SVC is expressed as a linear combination of $L$ approximate eigenvectors $\hat{\mathbf{E}}$ ($N \times L$). In the estimation step, we first apply (ii) a pre-compression, and the SVCs are compressed into $K^2(L+1)^2 + K + 1$ inner products that appear in step (a) in Section 4.2.1. Owning to this step, the computational cost for the likelihood evaluation is independent of $N$. Furthermore, the likelihood maximization is accelerated by (iii) a sequential estimation. The estimation in (iii) iterates processing of $O(L^3)$. Because the cost is independent of $N$, the estimation is fast even if $N$ is very large. Yet, the computational cost increases as $K$ growth. This outcome is because large $K$ increases the number of iterations to estimate $\boldsymbol{\theta}_1,... \boldsymbol{\theta}_K$ sequentially. The computational cost can be reduced by applying SVCs only on some focused explanatory variables, as explained before.

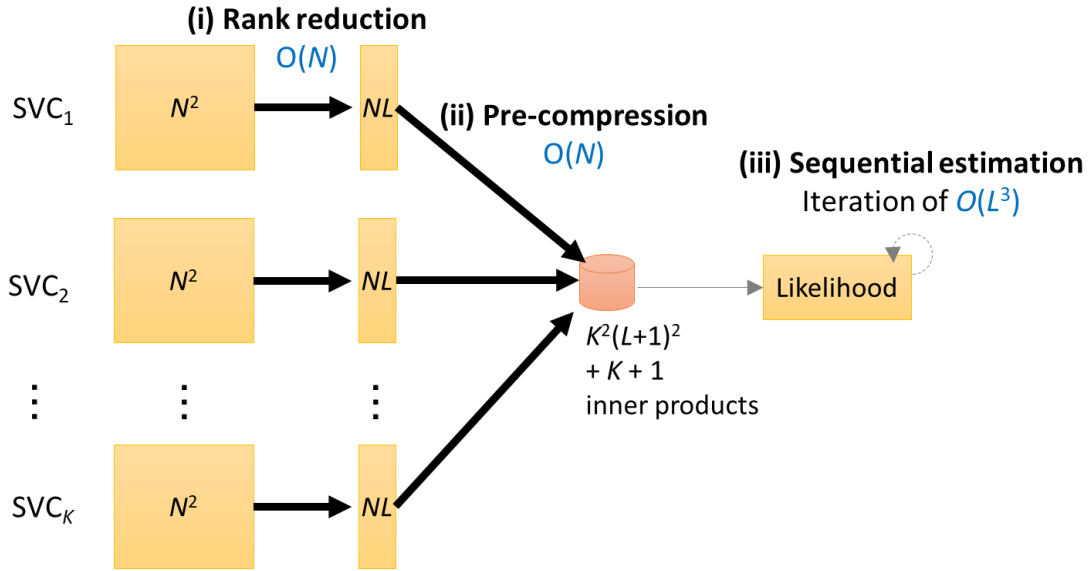

Figure 1: Overview of the proposed approach

## 5. Monte Carlo simulation experiment

This section compares our approach with existing approaches in terms of computational time and estimation accuracy. We deploy three simulation experiments. The first experiment compares our approach with the original M-SVC, GWR, and FB-GWR approaches assuming $N \leq 4,000$. The second simulation compares (relatively) fast approaches, including GWR and M-SVC, assuming $6,000 \leq N \leq 12,000$. The third experiment examines performance of our approach with larger samples ($20,000 \leq N \leq 100,000$). Throughout this section, we use a Mac Pro (3.5 GHz, 6-Core Intel Xeon E5 processor with 64 GB memory). R (version 3.6.2; https://cran.r-project.org/) is used for the model estimation. The package GWmodel (version 2-0.5; see Lu et al., 2014) is used to estimate GWR and FB-GWR models.

### 5.1. Simulation with relatively small samples ($N \leq 4,000$)

#### 5.1.1. Outline

Synthetic data $\{\mathbf{y}, \mathbf{x}_1,... \mathbf{x}_k\}$ are generated from

$$\mathbf{y} = \sum_{k=1}^{K} \mathbf{x}_k \circ \boldsymbol{\beta}_k + \boldsymbol{\varepsilon}, \qquad \mathbf{x}_k \sim N(\mathbf{0}, \mathbf{I}), \qquad \boldsymbol{\varepsilon} \sim N(\mathbf{0}, \sigma^2 \mathbf{I}), \qquad (23)$$

where $\boldsymbol{\beta}_k$ is generated with the following spatial moving average specification:

$$\boldsymbol{\beta}_k = \mathbf{1} + \bar{\mathbf{C}}\boldsymbol{\varepsilon}_k, \qquad \boldsymbol{\varepsilon}_k \sim N(\mathbf{0}, \mathbf{I}), \tag{24}$$

where $\bar{\mathbf{C}}$ is the row-standardize version of a proximity matrix whose $(i, j)$-th element equals exp(-$d(s_i, s_j)$). The spatial coordinates are drawn from two independent normal random variables.[2] The residual variance $\sigma^2$ is given by $0.3 Var\left[\sum_{k=1}^{K} \mathbf{x}_k \circ \boldsymbol{\beta}_k\right]$; in this case, the $R$-squared of this model is always $Var\left[\sum_{k=1}^{K} \mathbf{x}_k \circ \boldsymbol{\beta}_k\right] / 1.3 Var\left[\sum_{k=1}^{K} \mathbf{x}_k \circ \boldsymbol{\beta}_k\right] = 1/1.3 \approx$ 0.77. Because our preliminary analysis shows that the influence of $\sigma^2$ on the estimation accuracy is small, it is fixed hereafter.

This section compares SVC modeling approaches for 32 cases comprising of $N$ ∈ {500, 1,000, 1,500, 2,000, 2,500, 3,000, 3,500, 4,000} and $K$ ∈ {2, 4, 6, 8}. In each case, estimations involve 200 iterations.

Estimation accuracy is evaluated using the root mean squared error (RMSE) and bias, whose formulas are given by

$$RMSE[\hat{\beta}_k(s_i)] = \sqrt{\frac{1}{NP} \sum_{i=1}^{N} \sum_{p=1}^{200} \left(\beta_k(s_i) - \hat{\beta}_k(s_i)\right)^2}, \tag{25}$$

$$Bias[\hat{\beta}_k(s_i)] = \frac{1}{NP} \sum_{i=1}^{N} \sum_{p=1}^{200} \left(\beta_k(s_i) - \hat{\beta}_k(s_i)\right). \tag{26}$$

---

[2] This assumption implies fewer samples near periphery areas. It is a typical case in regional science, in which data typically have fewer samples in suburban areas of cities.

### 5.1.2. Result

This section fist quantifies influence from the following three approximations: (i) the eigen-approximation, (ii) the pre-compression, and (iii) the sequential estimation.

Regarding (i), the original M-SVC model (M-SVC) and our M-SVC with the eigen-approximation (M-SVC$_{(i)}$) are compared. Figure 2 portrays RMSEs and biases. Solid and dotted lines represent the results from M-SVC$_{(i)}$ and the original M-SVC, respectively. These figures show that their RMSEs and biases are almost the same. Figure 3 shows the mean correlation coefficients between SVCs estimated with the two approaches. The smallest value of 0.985 occurs when $N = 500$ and $K = 8$. The correlation coefficients increase as $N$ increases, and the values are greater than 0.997 in all cases when $N = 4,000$. These results confirm that the error due to (i) the Moran's eigenvector approximation is quite small.

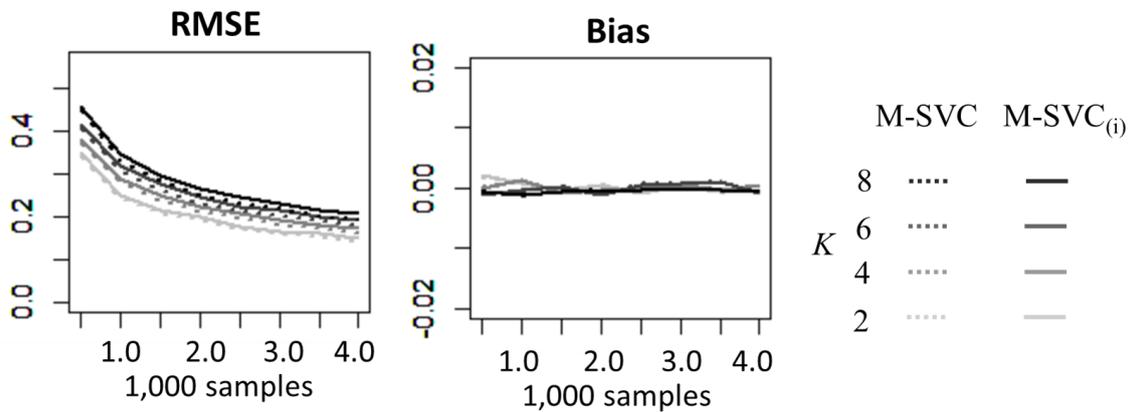

Figure 2: RMSEs and mean biases of the M-SVC and M-SVC$_{(i)}$ estimates ($N \leq 4,000$)

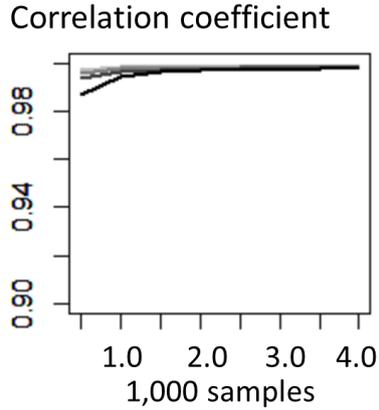

Figure 3: Mean correlation coefficients between the M-SVC and M-SVC$_{(i)}$ estimates ($N \leq 4,000$)

The computational gain by the approximation is substantial. Because the computational cost for the eigen-decomposition increases rapidly—it is of the order of $N^3$—the original M-SVC approach is not feasible for large samples, say $N > 10,000$. The cost for our eigen-approximation, which is based on the Nystrom extension (see Murakami and Griffith, 2018), increases only linearly with respect to $N$; it is available for large samples. In summary, the eigen-approximation considerably accelerates computation with a small approximation error.

Regarding (ii), the pre-compression, and (iii), the sequential estimation, we compare M-SVC$_{(i)}$, M-SVC$_{(ii)}$ using (i)-(ii), and M-SVC $_{(iii)}$ using (i)-(iii). Because of the computational burden here, the simulations employ only 20 iterations.

Figures 4 and 5 compare bias and RMSE between M-SVC$_{(i)}$/M-SVC$_{(ii)}$ and M-SVC$_{(iii)}$.[3] Each approach indicates similar bias and RMSE values. Bias is small across cases. RMSE decreases as $N$ increases and $K$ decreases. Figure 6 plots the minimum (i.e., the worst) correlation coefficients between the SVC estimates of M-SVC$_{(i)}$/M-SVC$_{(ii)}$ and M-SVC$_{(iii)}$. The minimum value results in 0.958 when $N = 500$ and $K = 8$. When $1,000 \leq N$, the value is always greater than 0.990. It is verified that the error introduced by (iii), the sequential estimation, is quite small, whereas (ii) does not produce any error.

Besides, the correlation coefficients summarized in Figures 3 and 6 reveal that the SVCs estimated with the original M-SVC and our M-SVC $_{(iii)}$ are almost the same. This result implies that the accuracy of the M-SVC approach, which is better than the standard GWR and the same as FB-GWR, holds for M-SVC$_{(iii)}$ (see, Murakami et al., 2018).

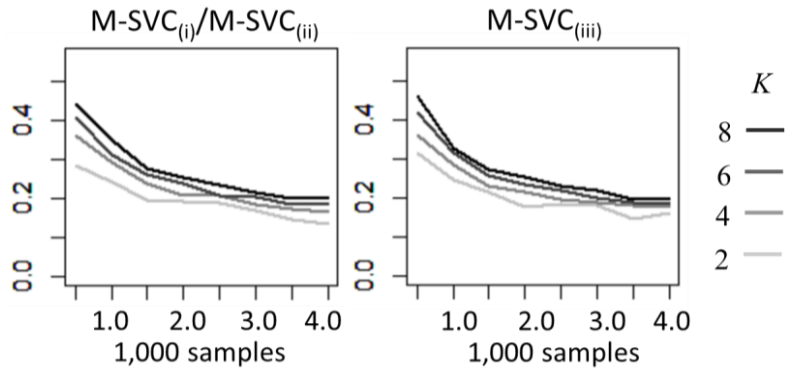

Figure 4: RMSEs of the M-SVC$_{(i)}$/ M-SVC$_{(ii)}$ and M-SVC$_{(iii)}$ estimates ($N \leq 4,000$)

---

[3] Because (ii), the pre-compression, does not introduce errors, estimates of M-SVC$_{(i)}$ and M-SVC$_{(ii)}$ are exactly identical.

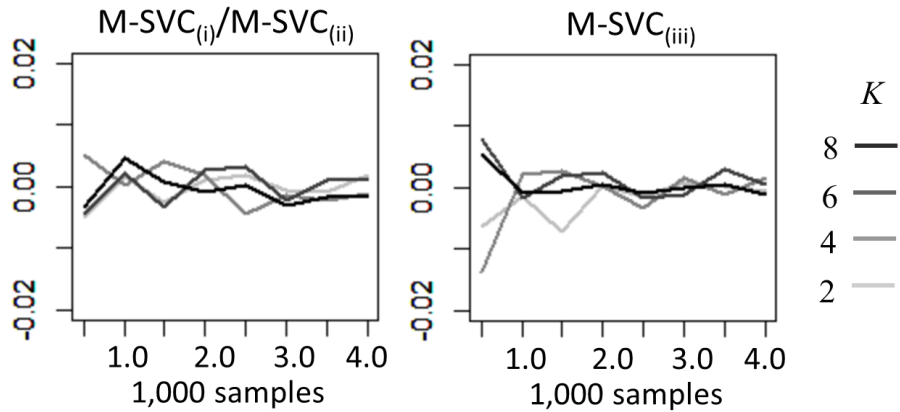

Figure 5: Mean biases of the M-SVC$_{(i)}$/ M-SVC$_{(ii)}$ and M-SVC$_{(iii)}$ estimates ($N \leq$ 4,000)

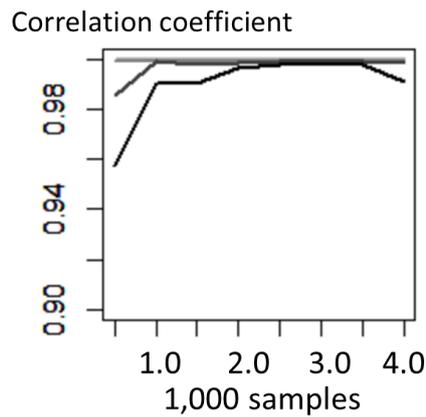

Figure 6: Mean correlation coefficients between the M-SVC$_{(i)}$/ M-SVC$_{(ii)}$ and M-SVCii$_{(i)}$ estimates ($N \leq$ 4,000)

Figure 7 compares computational times of M-SVC$_{(i)}$, M-SVC$_{(ii)}$, M-SVC$_{(iii)}$, GWR, and FB-GWR. M-SVC$_{(i)}$ is the slowest when $K \geq 6$, whereas FB-GWR is the slowest when $K \leq 4$. Although M-SVC$_{(ii)}$ is faster than M-SVC$_{(i)}$, it is still slow. For example, when $N = 4{,}000$ and $K = 8$, M-SVC$_{(ii)}$ took 2,903 seconds and M-SVC$_{(i)}$ took 5,270 second. These results indicate that (i), the eigen-approximation, and (ii), the pre-compression, are not sufficient for fast SVC estimation. However, M-SVC$_{(iii)}$ is considerably faster than either M-SVC$_{(i)}$ or M-SVC$_{(ii)}$. When $N = 4{,}000$ and $K = 8$, its estimation took 453.6 seconds on average. Yet, it is slower than GWR, which took only 195.1 seconds on average in the same case.

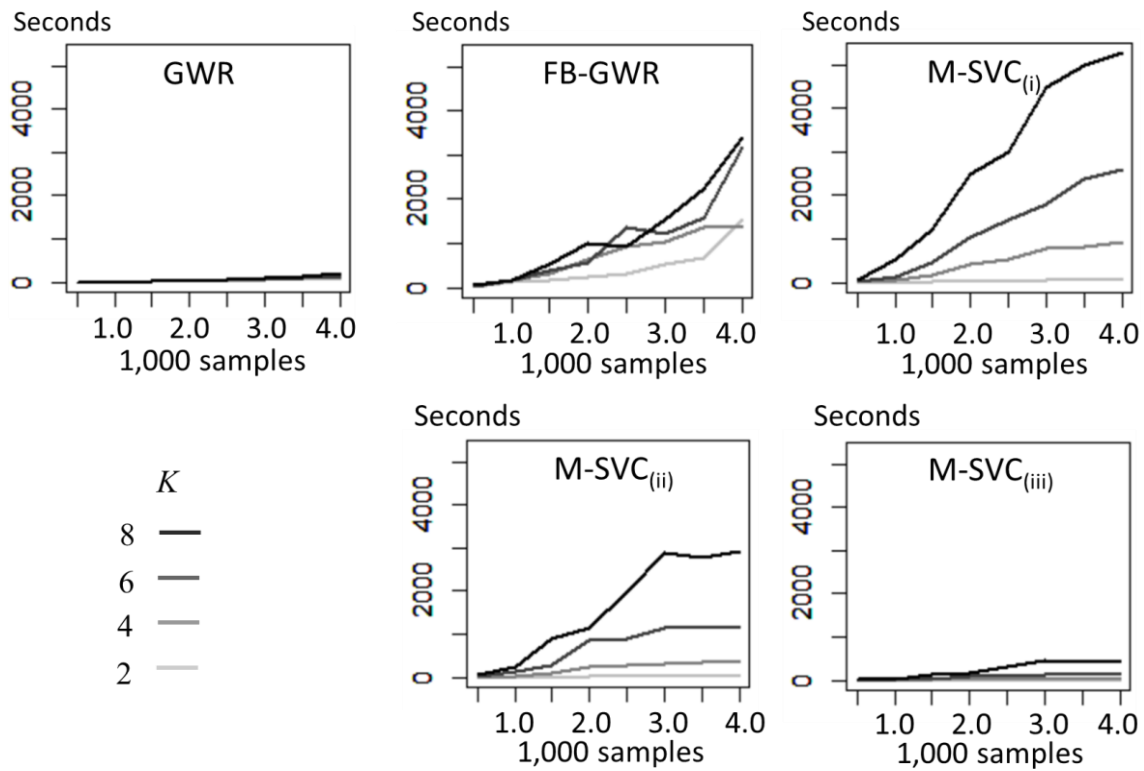

Figure 7: Computational time ($N \leq 4{,}000$)

## 5.2. Simulation with large samples (5,000 $\leq N \leq$ 100,000)

### 5.2.1. Outline

This section compares estimation accuracy and computational time for M-SVC$_{(iii)}$, which was found to be accurate and relatively fast when $N$ is small, with those for GWR using large samples. FB-GWR is not compared because it is computationally too expensive for large $N$. The large synthetic data are generated with Eqs. (27) and (28):

$$\mathbf{y} = \sum_{k=1}^{K} \mathbf{x}_k \circ \boldsymbol{\beta}_k + \boldsymbol{\varepsilon}, \qquad \boldsymbol{\varepsilon} \sim N(\mathbf{0}, \sigma^2 \mathbf{I}), \tag{27}$$

$$\boldsymbol{\beta}_k = \mathbf{1} + \widehat{\mathbf{E}}_+ \boldsymbol{\gamma}_k, \qquad \boldsymbol{\gamma}_k \sim N\big(\mathbf{0}, \widehat{\boldsymbol{\Lambda}}_+^{\alpha_k}\big). \tag{28}$$

where $\mathbf{x}_k$ is randomly generated from $N(\mathbf{0}, \mathbf{I})$ if $k > 1$, and $\mathbf{x}_k = \mathbf{1}$ if $k = 1$. To generate SVCs within a reasonable time frame, $\boldsymbol{\beta}_k$ is generated from $\widehat{\mathbf{E}}_+ \boldsymbol{\gamma}_k$, where $\widehat{\mathbf{E}}_+$ is a $N \times L_+$ matrix of Moran eigenvectors corresponding to positive eigenvalues, and $\widehat{\boldsymbol{\Lambda}}_+$ is a $L_+ \times L_+$ diagonal matrix of the positive eigenvalues (i.e., $\hat{\lambda}_l > 0$). $\widehat{\mathbf{E}}_+$ and $\widehat{\boldsymbol{\Lambda}}_+$ are generated as follows: (a) eigen-pairs for the $L$ knots are generated from Eqs. (12) and (13) using $\mathbf{C}_L$, which is the 2,000 × 2,000 matrix whose $(i, j)$-th element equals exp(-$d(s_i, s_j)$); and, (b) eigen-pairs corresponding to $\hat{\lambda} \leq 0$ are eliminated. Thus, $\boldsymbol{\beta}_k$ is given by a positively dependent spatial process using at most 2,000 eigenvectors corresponding to $\hat{\lambda}_l > 0$. Our preliminary analysis showed that the number of eigenvalues satisfying $\hat{\lambda} > 0$ is almost always less than 2,000 even when $N$ = 100,000. In other worse, the assumption of a maximum of 2,000 eigenvectors is reasonable considering the cost for computation. Note that the 2,000 eigenvectors are used only for data generation; as before, at most, 200 eigenvectors are used for the M-SVC model estimation.

The $\alpha_k$ parameter determines the spatial scale of $\boldsymbol{\beta}_k$. To test the influence of scale on estimation accuracy, assume $\alpha_k = 2$, which represents a large scale, for the first $K/2$ SVCs, and $\alpha_k = 0.5$, which represents a small scale, for the remaining $K/2$ SVCs.

Hereafter, Section 5.2.2 compares M-SVC$_{(iii)}$ and GWR assuming $N \leq 12,000$. Section 5.2.3 applies M-SVC$_{(iii)}$ to larger samples. In both sections, estimation involves 200 iterations in each case.

### 5.2.2.  Result ($N \leq 12,000$)

This section compares M-SVC$_{(iii)}$ to GWR assuming $N \in \{6,000, 9,000, 12,000\}$ and $K \in \{2, 4, 6, 8\}$.

Figure 8 portrays the mean bias for the large-scale SVCs (left) and the small-scale SVCs (right). This result shows that the biases are quite small irrespective of scale. Figure 9 displays the mean RMSEs. This result shows that the RMSE of M-SVC$_{(iii)}$ is considerably smaller than that for GWR if SVCs have a large-scale map pattern. Actually, as illustrated by Figure 10, M-SVC$_{(iii)}$ estimates large-scale SVCs almost perfectly, whereas GWR estimates have somewhat noisy patterns relative to the true values. The result is reasonable because M-SVC$_{(iii)}$ uses the first $L$ eigenvectors describing the $L$ largest-scale map patterns explained by the Moran coefficient (see Griffith, 2003).

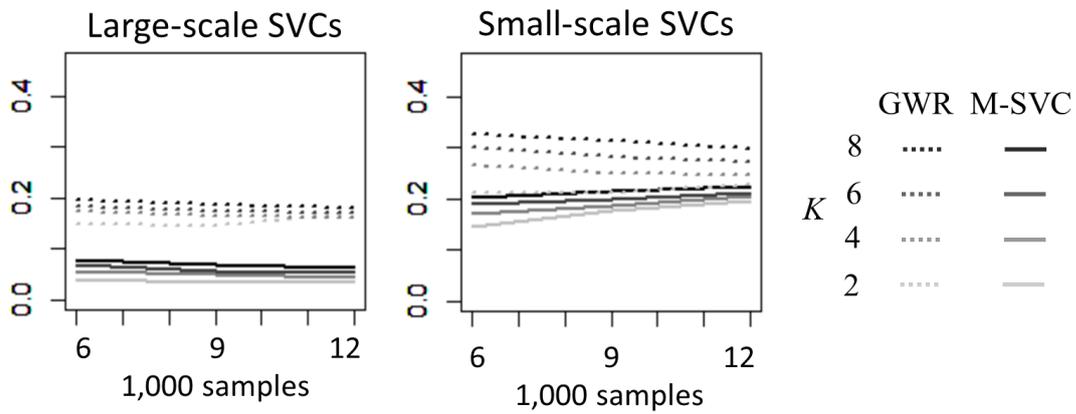

Figure 8: RMSEs of the GWR and M-SVC$_{(iii)}$ estimates (6,000 $\leq N \leq$ 12,000)

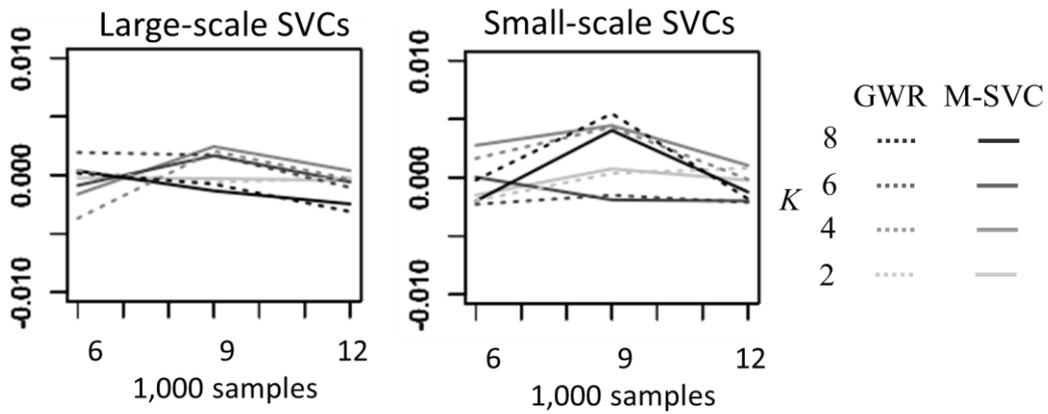

Figure 9: Mean bias of the GWR and M-SVC$_{(iii)}$ estimates (6,000 $\leq N \leq$ 12,000)

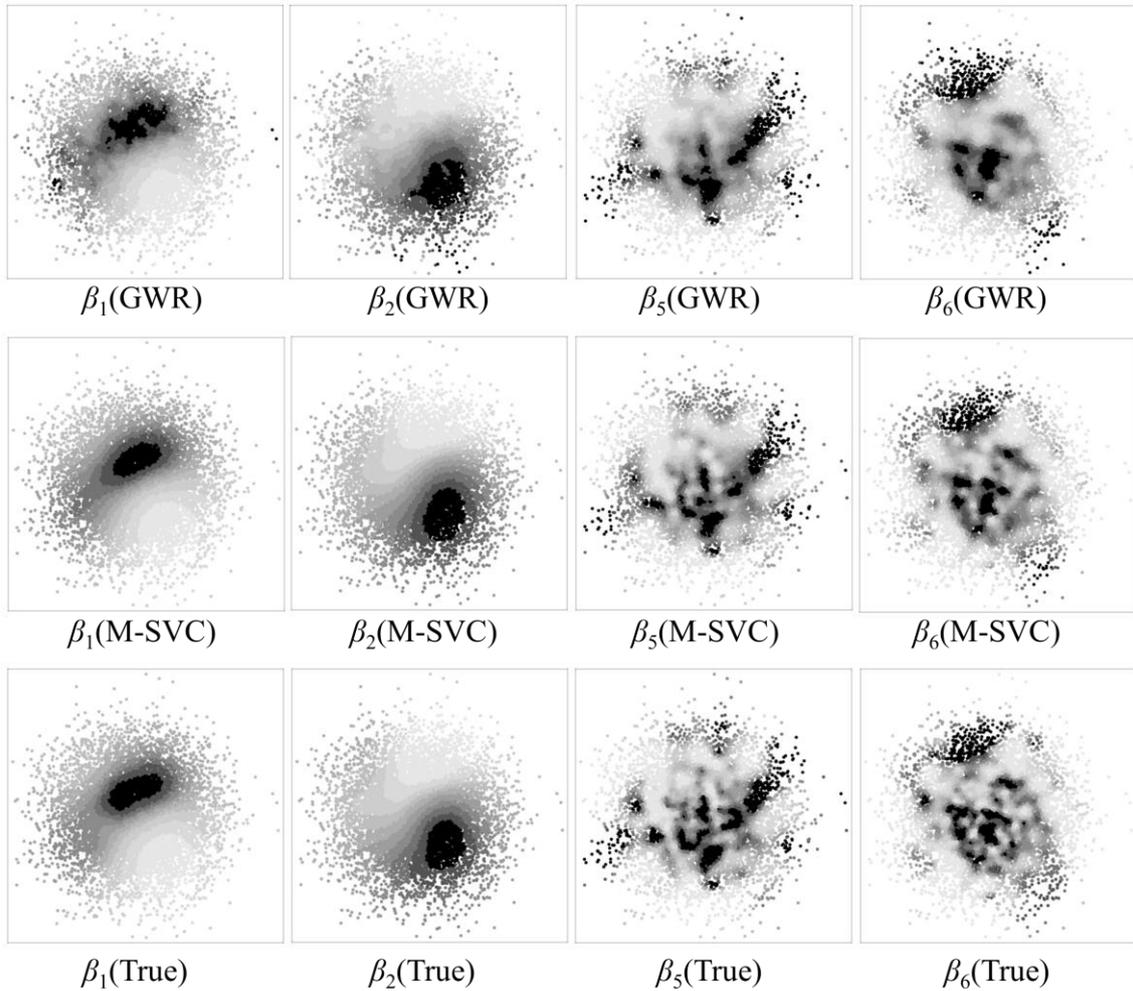

$\beta_1$(GWR)    $\beta_2$(GWR)    $\beta_5$(GWR)    $\beta_6$(GWR)

$\beta_1$(M-SVC)    $\beta_2$(M-SVC)    $\beta_5$(M-SVC)    $\beta_6$(M-SVC)

$\beta_1$(True)    $\beta_2$(True)    $\beta_5$(True)    $\beta_6$(True)

Figure 10: Plot of SVCs estimated in the first iteration ($N$ = 9,000)

Despite that M-SVC(iii) considers only the $L$ large-scale eigenvectors, accuracy of M-SVC(iii) estimates for small-scale $\boldsymbol{\beta}_k$ is still better than estimates from GWR. This might be because M-SVC(iii) estimates the scale $\alpha_k$ for each $\boldsymbol{\beta}_k$, whereas GWR estimates a common scale/bandwidth for all $\boldsymbol{\beta}_k$s. One finding is that M-SVC(iii) declines in accuracy as $N$ increases. This is because M-SVC(iii) limits $L$ to no more than 200 (see Section 4.1), and these 200 eigenvectors explain large-scale variations. Overcoming this limitation is an important future task. Still, true small-scale SVCs and those estimated by M-SVC$_{(iii)}$ are visually quite similar (Figure 10), whereas GWR estimates tends to be blurrier than the true distribution, probably due to its assumption of the same scale across $\beta_k$s.

Thus, the accuracy of M-SVC$_{(iii)}$ is verified as least when $N \leq 12{,}000$. A comparison of computational time is summarized in the next subsection.

### 5.2.3. Result ($N \leq 100{,}000$)

The previous subsection shows that accuracy of M-SVC$_{(iii)}$ for small-scale $\beta_k$ decreases as $N$ increases. To clarify whether M-SVC$_{(iii)}$ estimates remain accurate and computationally tractable for larger samples, we performed another simulations with $N \in \{20{,}000, 40{,}000, 70{,}000, 100{,}000\}$ and $K \in \{2, 4, 6, 8\}$, assuming the same data generating process used for Section 5.2.2.

Figure 11 (left), which plots biases, shows that bias is small, especially when $20{,}000 \leq N$. Figure 11 (right), which summarizes RMSEs, demonstrates that RMSE for large-scale SVCs with $\alpha_k = 2.0$ declines as $N$ increases, owning to the large samples that makes identification of SVCs easy. As shown in Figure 12, the correlation coefficient between large-scale SVC estimates and the true values are very close to 1.00, which confirms the estimation accuracy. Actually, as illustrated in Figure 13, true large-scale SVCs and the estimates are visually indistinguishable.

Regarding the small-scale SVCs with $\alpha_k = 0.5$, estimation accuracy decreases as $N$ increases because of the fixed $L$, as previously explained. However, the RMSE increase is quite small when $40,000 \leq N$. As a result, the mean correlation coefficient between the true and estimated SVCs is approximately 0.83, even when $N = 100,000$. Owning to the large size of $N$, estimation accuracy conceivably is almost the same even if the number of SVCs, $K$, is increased to 8. Figure 13 visually demonstrates that M-SVC successfully estimates small-scale patterns that are similar to the true values. These outcomes verified that our approach estimates both large- and small-scale SVCs with reasonable accuracy, even if $N$ is large. However, the estimated SVCs are slightly blurrier than the true values. Mitigating this problem while saving computational time is an important next step.

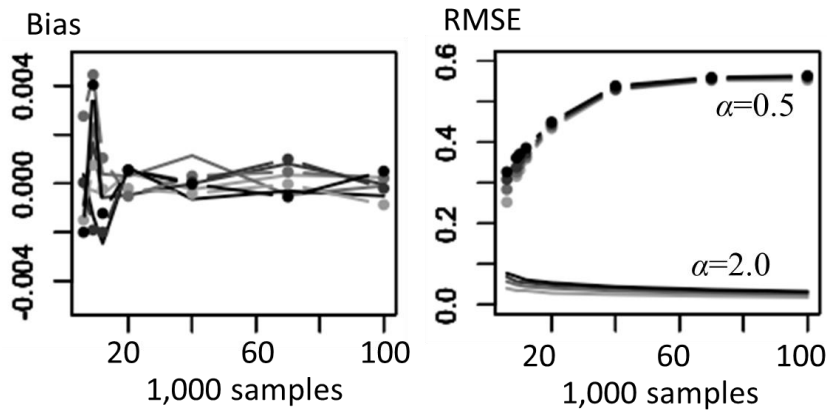

Figure 11: Mean biases and RMSEs of the M-SVC$_{(iii)}$ estimates. $\alpha$ =2.0 implies large-scale SVCs, and $\alpha$ =0.5 implies small-scale SVCs ($N \leq 100,000$).

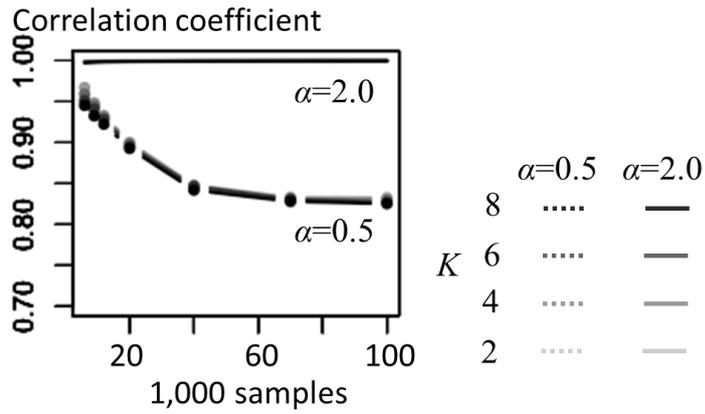

Figure 12: Correlation coefficient between the true SVCs and the M-SVC$_{(iii)}$ estimates.

$\alpha$ =2.0 implies large-scale SVCs, and $\alpha$ =0.5 implies small-scale SVCs ($N \leq 100,000$).

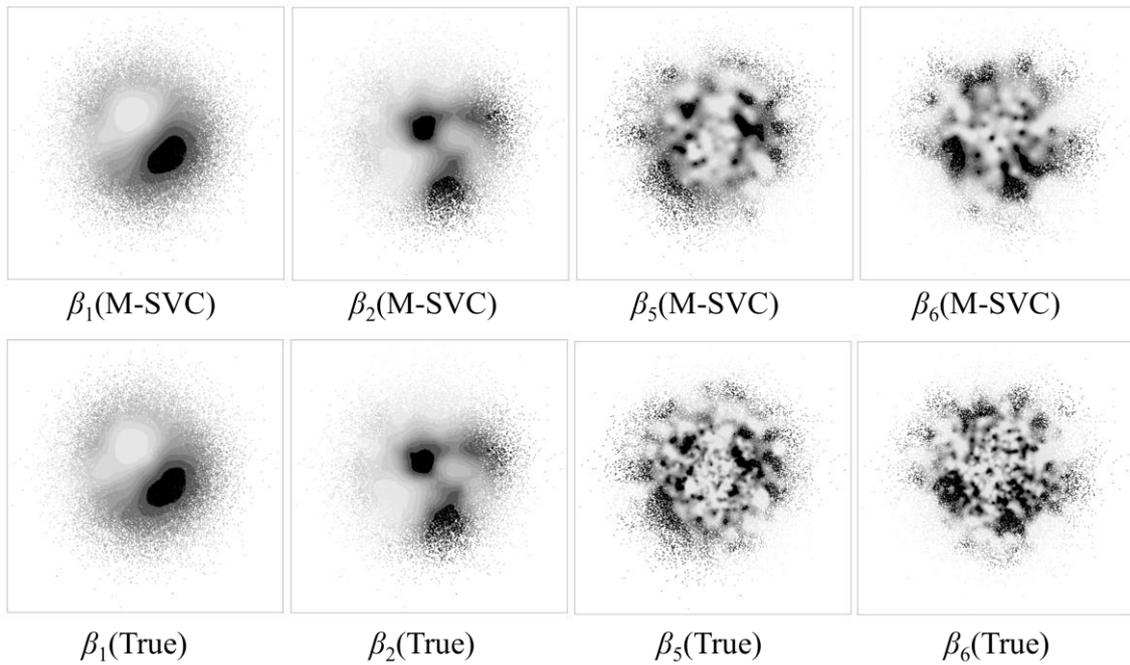

Figure 13: Plot of SVCs estimated in the first iteration ($N$ = 70,000)

Finally, Figure 14 summarizes computational time for GWR and M-SVC$_{(iii)}$. Although GWR is fast for small samples, as shown in Section 5.1.2, its computational time rapidly increases as $N$ grows because of the bandwidth optimization involved.

In contrast, an increase in computational time is surprisingly slow in the case of M-SVC$_{(iii)}$. For example, when $K = 8$, if $N$ is increased from 9,000 to 100,000, the mean computational time increases from 534 seconds to 835 seconds. This computational efficiency is because our approach processes $N$ samples in (i), the eigen-approximation, and (ii), the pre-compression steps, but these steps are required only one time, and their both of computational costs are $O(N)$. This is the reason why the computational time increase is linear and slow. Although these linear increases explain the cost for (i) and (ii), the computational time that is needed independent of $N$ explains the cost for (iii), the sequential estimation. Based on Figure 14, the computational time for (iii) is around 500 seconds across cases when $K = 8$. Thus, the computational cost for (iii), the estimation, is also quite small despite this step needing to numerically estimate 16 parameters $\{\boldsymbol{\theta}_1,...\boldsymbol{\theta}_8\}$, whereas GWR numerically optimizes only 1 bandwidth parameter.

Therefore, M-SVC$_{(iii)}$ is found to be accurate enough and extremely fast.

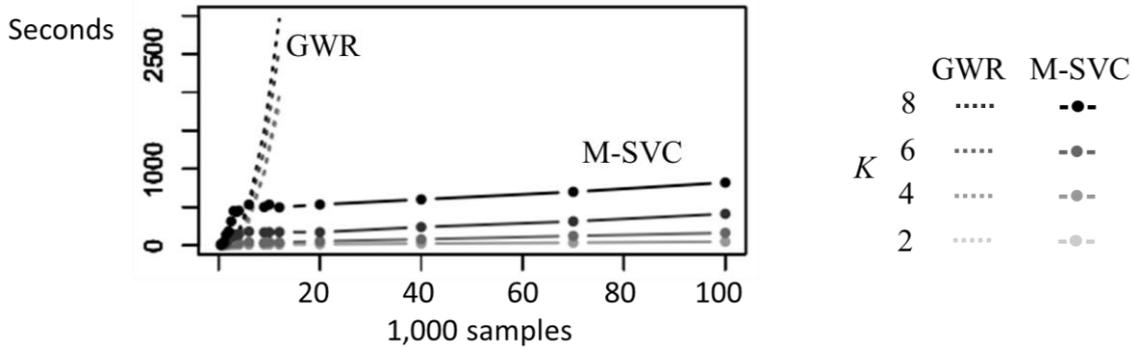

Figure 14: Computational time ($N \leq 100{,}000$)

## 6. An empirical study

As an illustration, GWR and M-SVC(iii) are applied to the officially assessed residential land price data in the Tokyo metropolitan area in 2010 ($N = 7{,}679$). The response variable is logged land prices [JPY/m²]. Covariates are distance to the nearest railway station (Station_d) [*km*], distance to the Tokyo railway station (Tokyo_d) [*km*], share of green area in 1km grids (Green), and anticipated flooding depth (Flood) [m]. These data were obtained from the National Land Numerical Information download service (http://nlftp.mlit.go.jp/ksj-e/index.html), provided by MLIT, Japan.

Figure 15 plots estimated coefficients (except for spatially varying intercepts). GWR took 721 seconds for its estimation, while the M-SVC approach took 241 seconds. The SVCs estimated with GWR have similar scales for their spatial patterns. In contrast, SVCs estimated with our M-SVC approach have considerably different scales for their spatial patterns; specifically, the SVCs for Tokyo_d have a large-scale spatial pattern, the SVCs for Forest have a moderate-scale pattern, and those for Station_d and Flood have small-scale patterns. The SVCs for Tokyo_d have large negative values in central Tokyo and the northern part of the metropolitan area; this suggests that accessibility to Tokyo encourages urbanization, especially to the north. Coefficients for Station_d reveal that station access has a stronger negative relationship with land price in suburban residential areas. Because Tokyo_d and Station_d can be viewed as global and local-measures of accessibility, large- and small-scale spatial patterns in these SVC estimates are intuitively reasonable. SVCs for Forest show that green areas has strong negative relationships in suburban non-urban areas. Finally, SVCs for Flood have a stronger negative relationship around the Arakawa river, which is a principal river in this area. It suggests that flood risk is appropriately reflected in land prices in the Arakawa coastal areas. In contrast, GWR estimates seem less interpretable because of their same scale across SVCs.

In conclusion, the M-SVC approach provides reasonable estimates of multiscale SVCs.

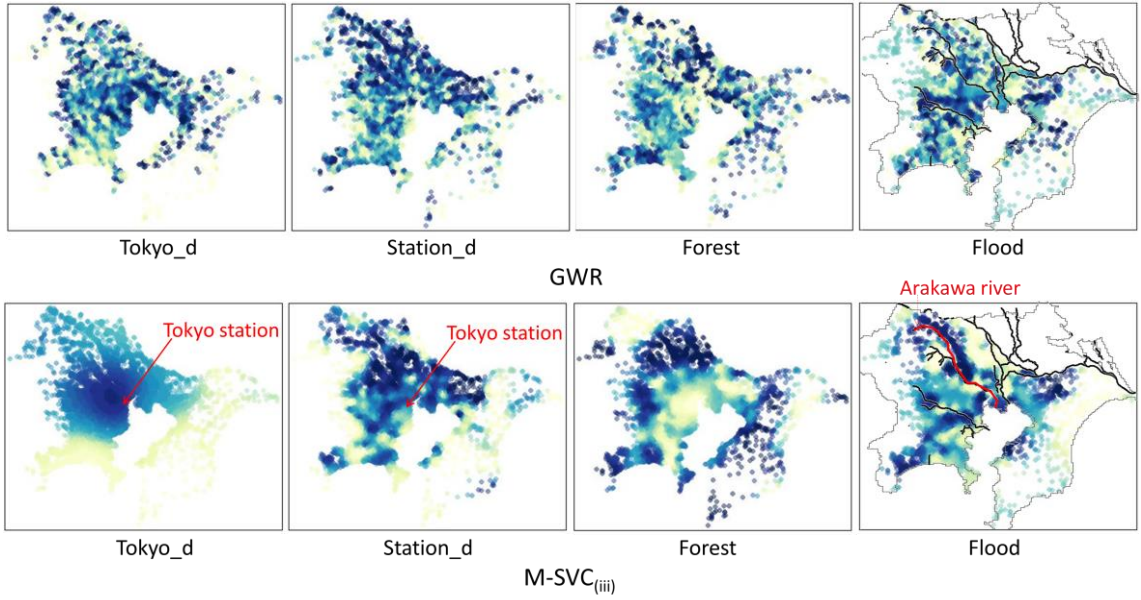

Figure 15: Estimated SVCs. For the panel on Flood, principal revers are displayed.

# 7. Future directions

While large spatial data modeling is a recent hot topic, related discussions are quite limited when it comes to SVC modeling. Given this background, this study develops a fast M-SVC approach that estimates multiscale SVCs. We achieve the computational efficiency using (i) rank reduction, (ii) pre-compression, and (iii) sequential estimation. Monte Carlo simulation experiments confirm the estimation accuracy and computational efficiency of our approach. Our approach estimates SVCs in relatively little time, without applying any parallel computation. In other words, a high performance computing environment is not needed for our approach.

Nevertheless, step (i) is easily parallelized, similar to the other Nystrom extension-based eigen-approximation approaches (see Li et al., 2015). The same holds for step (ii), which simply calculates inner-products. Although parallelizing the sequential estimation step (iii) is not straightforward, its computational cost is independent of $N$. Thus, parallelization of our approach allows ever faster estimation of multiscale SVCs, and extends results summarized here to millions of samples.

One drawback of our approach is its incapability of modeling fine-scale spatial variations, as illustrated in Figure 13; we consider at most 200 eigenvectors corresponding to large eigenvalues, which implies ignoring the other eigenvectors that explain small-scale variations. Such a limitation in low rank spatial modeling is reported in geostatistsics (Stein, 2014). Addressing this problem is an important future research topic. Fortunately, a number of multiscale approaches have been proposed for this problem (e.g., Sang and Huang, 2012; Nychika et al., 2015).

Local approaches are more suitable than rank reduction approaches, including ours, to capture fine-scale variations (Stein, 2014). In this respect, acceleration of GWR is a sensible way to estimate fine-scale SVCs with large samples. Considering the similar estimation accuracy for the M-SVC and FB-GWR approaches (see Murakami et al., 2018), acceleration of FB-GWR, GWR with parametric-specific distance matrices (PSDM-GWR; Lu et al., 2017), and other extended GWRs, also is a promising way to develop fast and flexible SVC modeling approaches. Studies about fast GWR include Harris et al. (2010B), Tran et al. (2016), and Lu et al. (2017; 2018).

Although we develop a fast approach for SVC estimation, the same approach potentially is available to other mixed effects model, which can be written as Eq. (7); they include spatial (or non-spatial) additive models, and multilevel models, among others. Extension of our fast estimation framework to a wide variety of spatial regression models would be valuable in the era of big spatial data.

The M-SVC modeling approach is implemented in an R packages "spmoran" (https://cran.r-project.org/web/packages/spmoran/index.html; see, Murakami, 2018) case study examining violent crime rates and their relationships to alcohol outlets and illegal drug arrests. J. Geog. Sci. 11 (1), 1–22.

- Yang, W., 2014. An extension of geographically weighted regression with flexible bandwidths. PhD Thesis. University of St Andrews.

- Zhang, K., Kwok, J.T., 2010. Clustered Nyström method for large scale manifold learning and dimension reduction. IEEE Trans. Neural Netw. 21 (10), 1576–1587.

## Appendix.1 Deviation of Eq.(21)

For simplicity, $\mathbf{V}(\boldsymbol{\theta}_k)$ is written as $\mathbf{V}_k$. The matrix $\mathbf{P} =$

$$\begin{bmatrix} \mathbf{M}_{0,0} & \mathbf{M}_{0,1}\mathbf{V}_1 & \cdots & \mathbf{M}_{0,K}\mathbf{V}_K \\ \mathbf{V}_1\mathbf{M}_{1,0} & \mathbf{V}_1\mathbf{M}_{1,1}\mathbf{V}_1 + \mathbf{I} & \cdots & \mathbf{V}_1\mathbf{M}_{1,K}\mathbf{V}_K \\ \vdots & \vdots & \ddots & \vdots \\ \mathbf{V}_K\mathbf{M}_{K,0} & \mathbf{V}_K\mathbf{M}_{K,1}\mathbf{V}_1 & \cdots & \mathbf{V}_K\mathbf{M}_{K,K}\mathbf{V}_K + \mathbf{I} \end{bmatrix}$$ has the following expression:

$$\mathbf{P} = \begin{bmatrix} \widetilde{\mathbf{V}}_{-K} & \mathbf{O} \\ \mathbf{O} & \mathbf{V}_K \end{bmatrix} \begin{bmatrix} \widetilde{\mathbf{M}}_{-K,-K} + \widetilde{\mathbf{V}}_{-K}^{-2} & \widetilde{\mathbf{M}}_{-K,K} \\ \widetilde{\mathbf{M}}_{K,-K} & \mathbf{M}_{K,K} + \mathbf{V}_K^{-2} \end{bmatrix} \begin{bmatrix} \widetilde{\mathbf{V}}_{-K} & \mathbf{O} \\ \mathbf{O} & \mathbf{V}_K \end{bmatrix}, \quad (A1)$$

where $\widetilde{\mathbf{V}}_{-K} = \begin{bmatrix} \mathbf{I} & & & \\ & \mathbf{V}_1 & & \\ & & \ddots & \\ & & & \mathbf{V}_{K-1} \end{bmatrix}, \widetilde{\mathbf{M}}_{-K,-K} = \begin{bmatrix} \mathbf{M}_{0,0} & \mathbf{M}_{0,1} & \cdots & \mathbf{M}_{0,K-1} \\ \mathbf{M}_{1,0} & \mathbf{M}_{1,1} & \cdots & \mathbf{M}_{1,K-1} \\ \vdots & \vdots & \ddots & \vdots \\ \mathbf{M}_{K-1,0} & \mathbf{M}_{K-1,1} & \cdots & \mathbf{M}_{K-1,K-1} \end{bmatrix}$, and

$\widetilde{\mathbf{M}}_{K,-K} = \begin{bmatrix} \mathbf{M}_{K,0} & \mathbf{M}_{K,1} & \cdots & \mathbf{M}_{K,K-1} \end{bmatrix}$.

The inverse of $\mathbf{P}$ is expressed using Eq.(A1) as

$$\mathbf{P}^{-1} = \begin{bmatrix} \widetilde{\mathbf{V}}_{-K}^{-1} & \mathbf{O} \\ \mathbf{O} & \mathbf{V}_K^{-1} \end{bmatrix} \begin{bmatrix} \widetilde{\mathbf{M}}_{-K,-K} + \widetilde{\mathbf{V}}_{-K}^{-2} & \widetilde{\mathbf{M}}_{-K,K} \\ \widetilde{\mathbf{M}}_{K,-K} & \mathbf{M}_{K,K} + \mathbf{V}_K^{-2} \end{bmatrix}^{-1} \begin{bmatrix} \widetilde{\mathbf{V}}_{-K}^{-1} & \mathbf{O} \\ \mathbf{O} & \mathbf{V}_K^{-1} \end{bmatrix}. \quad (A2)$$

Note that

$$\begin{bmatrix} \widetilde{\mathbf{M}}_{-K,-K} + \widetilde{\mathbf{V}}_{-K}^{-2} & \widetilde{\mathbf{M}}_{-K,K} \\ \widetilde{\mathbf{M}}_{K,-K} & \mathbf{M}_{K,K} + \mathbf{V}_K^{-2} \end{bmatrix} = \begin{bmatrix} \widetilde{\mathbf{M}}_{-K,-K} + \widetilde{\mathbf{V}}_{-K}^{-2} & \widetilde{\mathbf{M}}_{-K,K} \\ \widetilde{\mathbf{M}}_{K,-K} & \mathbf{M}_{K,K} \end{bmatrix} + \begin{bmatrix} \mathbf{O} \\ \mathbf{I} \end{bmatrix} \mathbf{V}_K^{-2} [\mathbf{O} \quad \mathbf{I}]. \quad \text{(A3)}$$

Suppose that $\mathbf{Q} = \begin{bmatrix} \widetilde{\mathbf{M}}_{-K,-K} + \widetilde{\mathbf{V}}_{-K}^{-2} & \widetilde{\mathbf{M}}_{-K,K} \\ \widetilde{\mathbf{M}}_{K,-K} & \mathbf{M}_{K,K} \end{bmatrix}$ and $\mathbf{Q}^{-1} = \begin{bmatrix} \mathbf{Q}_{-K,-K}^* & \mathbf{Q}_{-K,K}^* \\ \mathbf{Q}_{K,-K}^* & \mathbf{Q}_{K,K}^* \end{bmatrix}$, Eq.(A3) is

inverted using the Woodbury matrix identity as follows

$$\begin{bmatrix} \widetilde{\mathbf{M}}_{-K,-K} + \widetilde{\mathbf{V}}_{-K}^{-2} & \widetilde{\mathbf{M}}_{-K,K} \\ \widetilde{\mathbf{M}}_{K,-K} & \mathbf{M}_{K,K} + \mathbf{V}_K^{-2} \end{bmatrix}^{-1}$$
$$= \mathbf{Q}^{-1} - \begin{bmatrix} \mathbf{Q}_{-K,K}^* \\ \mathbf{Q}_{K,K}^* \end{bmatrix} (\mathbf{V}_K^2 + \mathbf{Q}_{K,K}^*)^{-1} [\mathbf{Q}_{K,-K}^* \quad \mathbf{Q}_{K,K}^*]. \quad \text{(A4)}$$

Only the second term includes the unknown parameters $\boldsymbol{\theta}_k$ (in $\mathbf{V}_k$).

$\mathbf{P}^{-1}$ is expressed by substituting Eq.(A4) into Eq.(A2), as follows:

$$\mathbf{P}^{-1} = \begin{bmatrix} \widetilde{\mathbf{V}}_{-K}^{-1} & \mathbf{O} \\ \mathbf{O} & \mathbf{V}_K^{-1} \end{bmatrix} \begin{bmatrix} \mathbf{Q} - \begin{bmatrix} \mathbf{Q}_{-K,K}^* \\ \mathbf{Q}_{K,K}^* \end{bmatrix} (\mathbf{V}_K^2 + \mathbf{Q}_{K,K}^*)^{-1} [\mathbf{Q}_{K,-K}^* \quad \mathbf{Q}_{K,K}^*] \end{bmatrix} \begin{bmatrix} \widetilde{\mathbf{V}}_{-K}^{-1} & \mathbf{O} \\ \mathbf{O} & \mathbf{V}_K^{-1} \end{bmatrix}$$
$$= \begin{bmatrix} \widetilde{\mathbf{V}}_{-K}^{-1} & \mathbf{O} \\ \mathbf{O} & \mathbf{V}_K^{-1} \end{bmatrix} \mathbf{Q}^{-1} \begin{bmatrix} \widetilde{\mathbf{V}}_{-K}^{-1} & \mathbf{O} \\ \mathbf{O} & \mathbf{V}_K^{-1} \end{bmatrix} \quad \text{(A5)}$$
$$- \begin{bmatrix} \widetilde{\mathbf{V}}_{-K}^{-1} \mathbf{Q}_{-K,K}^* \\ \mathbf{V}_K^{-1} \mathbf{Q}_{K,K}^* \end{bmatrix} (\mathbf{V}_K^2 + \mathbf{Q}_{K,K}^*)^{-1} [\mathbf{Q}_{K,-K}^* \widetilde{\mathbf{V}}_{-K}^{-1} \quad \mathbf{Q}_{K,K}^* \mathbf{V}_K^{-1}].$$

By substituting Eq.(A5) and $\mathbf{m}_{-K} = [\mathbf{m'}_0 \quad \mathbf{m'}_1 \quad \cdots \quad \mathbf{m'}_{K-1}]'$, Eq.(19) is expanded as

follows:

$$\begin{bmatrix} \hat{\mathbf{b}} \\ \hat{\mathbf{u}}_1 \\ \vdots \\ \hat{\mathbf{u}}_K \end{bmatrix} = \begin{bmatrix} \widetilde{\mathbf{V}}_{-K}^{-1} & \mathbf{O} \\ \mathbf{O} & \mathbf{V}_K^{-1} \end{bmatrix} \mathbf{Q}^{-1} \begin{bmatrix} \widetilde{\mathbf{V}}_{-K}^{-1} & \mathbf{O} \\ \mathbf{O} & \mathbf{V}_K^{-1} \end{bmatrix} \begin{bmatrix} \widetilde{\mathbf{V}}_{-K} \mathbf{m}_{-K} \\ \mathbf{V}_K \mathbf{m}_K \end{bmatrix}$$

$$- \begin{bmatrix} \widetilde{\mathbf{V}}_{-K}^{-1} \mathbf{Q}_{-K,K}^* \\ \mathbf{V}_K^{-1} \mathbf{Q}_{K,K}^* \end{bmatrix} (\mathbf{V}_K^2 + \mathbf{Q}_{K,K}^*)^{-1} [\mathbf{Q}_{K,-K}^* \widetilde{\mathbf{V}}_{-K}^{-1} \quad \mathbf{Q}_{K,K}^* \mathbf{V}_K^{-1}] \begin{bmatrix} \widetilde{\mathbf{V}}_{-K} \mathbf{m}_{-K} \\ \mathbf{V}_K \mathbf{m}_K \end{bmatrix}, \quad \text{(A6)}$$

$$= \begin{bmatrix} \widetilde{\mathbf{V}}_{-K}^{-1} & \mathbf{O} \\ \mathbf{O} & \mathbf{V}_K^{-1} \end{bmatrix} \mathbf{Q}^{-1} \begin{bmatrix} \mathbf{m}_{-K} \\ \mathbf{m}_K \end{bmatrix}$$

$$- \begin{bmatrix} \widetilde{\mathbf{V}}_{-K}^{-1} \mathbf{Q}_{-K,K}^* \\ \mathbf{V}_K^{-1} \mathbf{Q}_{K,K}^* \end{bmatrix} (\mathbf{V}_K^2 + \mathbf{Q}_{K,K}^*)^{-1} [\mathbf{Q}_{K,-K}^* \mathbf{m}_{-K} + \mathbf{Q}_{K,K}^* \mathbf{m}_K],$$

which equals Eq.(21).

## Appendix.2 Deviation of the determinant Eq.(22)

The determinant of the $\mathbf{P}$ matrix is expanded as follows:

$$
\begin{aligned}
|\mathbf{P}| &= \begin{vmatrix} \widetilde{\mathbf{V}}_{-K} & \mathbf{O} \\ \mathbf{O} & \mathbf{V}_K \end{vmatrix}^2 \begin{vmatrix} \widetilde{\mathbf{M}}_{-K,-K} + \widetilde{\mathbf{V}}_{-K}^{-2} & \widetilde{\mathbf{M}}_{-K,K} \\ \widetilde{\mathbf{M}}_{K,-K} & \mathbf{M}_{K,K} + \mathbf{V}_K^{-2} \end{vmatrix}, \\
&= \left| \widetilde{\mathbf{V}}_{-K} \right|^2 |\mathbf{V}_K|^2 \left| \widetilde{\mathbf{M}}_{-K,-K} + \widetilde{\mathbf{V}}_{-K}^{-2} \right| \left| \mathbf{M}_{K,K} + \mathbf{V}_K^{-2} - \widetilde{\mathbf{M}}_{K,-K} (\widetilde{\mathbf{M}}_{-K,-K} + \widetilde{\mathbf{V}}_{-K}^{-2})^{-1} \widetilde{\mathbf{M}}_{-K,K} \right|.
\end{aligned}
\tag{A7}
$$

Here, a formula that $\begin{vmatrix} \mathbf{A} & \mathbf{B} \\ \mathbf{B}' & \mathbf{D} \end{vmatrix} = |\mathbf{A}||\mathbf{D} - \mathbf{B}'\mathbf{A}^{-1}\mathbf{B}|$ where $\mathbf{A}$ and $\mathbf{D}$ are square matrixes

and $\mathbf{B}$ is a matrix with appropriate size, is used (see, Silvester, 2000).